\documentclass{aa}  

\usepackage{graphicx}
\usepackage[hyperindex, breaklinks, colorlinks, linkcolor=blue, citecolor=blue]{hyperref}
\usepackage{txfonts}
\usepackage{calrsfs}
\usepackage{upgreek}
\usepackage{xcolor}
\DeclareMathAlphabet{\pazocal}{OMS}{zplm}{m}{n}

\usepackage{hyperref}
\begin{document}

   \title{The mass-to-flux ratio in molecular clouds. What are we really measuring?}


   \author{Aris Tritsis
          \inst{1}}

   \institute{Institute of Physics, Laboratory of Astrophysics, Ecole Polytechnique F\'ed\'erale de Lausanne (EPFL), \\ Observatoire de Sauverny, 1290, Versoix, Switzerland \\
              \email{aris.tritsis@epfl.ch}}
   \date{Received date; accepted date}
   \titlerunning{Ambipolar diffusion and the mass-to-flux ratio in turbulent clouds}
   \authorrunning{A. Tritsis}
 
  \abstract
   {The mass-to-magnetic flux ratio of molecular clouds is a parameter of central importance as it quantifies the dynamical significance of the magnetic field with respect to gravitational forces. Therefore, it can provide invaluable information on the fate of clouds, and the sites of star formation.}
   {Our objective was to study the accuracy with which we can measure the true mass-to-flux ratio in molecular clouds under various projection angles and identify systematic biases.}
   {We used a 3D nonideal magnetohydrodynamic chemo-dynamical simulation of a turbulent collapsing molecular cloud. We quantified the accuracy with which the mass-to-flux ratio is recovered under various projection angles and dynamical stages by analyzing the magnetic field - gas column density relation, and comparing the ``observed'' mass-to-flux ratio against the true values.}
   {We find that projection effects have a major effect on measurements of the mass-to-flux ratio. Zeeman measurements can overestimate the true mass-to-flux ratio of the cloud by more than an order of magnitude when the magnetic field primarily lies on the plane of the sky. Therefore, measurements of the mass-to-flux ratio based on Zeeman observations should be considered as upper limits. Mass-to-flux ratio estimates inferred from polarization observations do not provide a physically meaningful probe of the true mass-to-flux ratio and can lead to unphysical results as they fail to capture the underlying correlation between the magnetic field and column density.}
   {}


   \maketitle


\section{Introduction}\label{intro}

A key quantity used to characterize the relative importance of magnetic fields with respect to the self-gravity of molecular clouds is the mass-to-magnetic flux ratio (usually denoted as $\mu$). Defined as the mass enclosed by a magnetic flux tube divided by the respective flux, if $\mu$ is below a critical value \citep{1976ApJ...210..326M}, gravity cannot overcome the magnetic support (pressure and tension), and the cloud will never collapse under flux-freezing conditions.

While the definition of $\mu$ is straightforward in theory, its practical measurement in molecular clouds is far from trivial since, observationally, we are limited in what we can access. Mass can only be measured along the line of sight (LOS), while the strength of the magnetic field ($\rm{B}$) can be measured or estimated either along the LOS (via Zeeman observations) or in the plane of the sky (POS) via dust polarization. Additionally, flux is fundamentally a surface-integrated quantity, defined as the dot product of the magnetic field with the unit vector normal to a surface ($\vec{da}$) which in observations will always be aligned with the LOS.

For Zeeman observations, the mass-to-flux ratio will be $\mu_{\rm obs} = \rm{M} / (\vec{\rm{B_{LOS}}} \cdot \vec{da}) = N_{p} / \rm{B_{LOS}}$, where $\rm{N_{p}}$ is the total column density of the gas. However, this quantity is not necessarily representative of the true mass-to-flux ratio that governs the dynamics of the cloud. Consider, for example, a cloud of any shape where the magnetic field lies entirely within the POS. Zeeman measurements would return $\rm{B_{LOS}} = 0$, leading to an infinite $\mu_{\rm obs}$, implying no magnetic support, despite the fact that the relevant $\mu$ is finite. Even in a scenario where the inclination angle is favorable and aligned to the mean component of the magnetic field, Zeeman measurements of $\mu_{\rm obs}$ can still deviate from the true mass-to-flux ratio. Deformations in the flux tubes can result in assigning the wrong mass to the wrong flux, hence misrepresenting the underlying magnetization of the cloud (see the schematic in Fig. 1 of \citealt{2009MNRAS.400L..15M} and Fig. 6 of \citealt{2025A&A...700A.152T}). These deviations can be further amplified by the LOS inclination and, as a result, a wide range of $\mu_{\rm obs}$ values can arise solely from the orientation of distorted field lines relative to the observer. Relaxing the commonly adopted assumption of straight-parallel field lines therefore renders the observational limits on $\mu_{\rm obs}$ far less restrictive. 

Given these complexities, it is very challenging to measure the so-called differential mass-to-flux ratio \citep{1991ApJ...373..169M}, that is, to accurately characterize the distribution of mass in the different flux tubes of a cloud. Yet, this distribution has critical implications for the cloud's evolution, as clouds with the same total mass and magnetic flux can behave very differently depending on how the flux is concentrated within their internal structure \citep{1978prpl.conf..209M}. Therefore, while Zeeman measurements probe a physically meaningful component of the field to define a flux, they may still misrepresent the cloud's magnetization and may not capture the full picture.

In the case of polarization observations, one might be tempted to derive a mass-to-flux ratio as $\mu_{\rm obs} = \rm{N_{p}} / \rm{B_{POS}}$. However, such an expression cannot be formally obtained as $\vec{da}$ and $\vec{\rm{B_{POS}}}$ are always orthogonal, their dot product is zero, and the area element does not cancel out. In other words, $\vec{\rm{B_{POS}}}$ does not thread the plane of the sky, and, by definition, the magnetic flux through it is zero, regardless of the cloud’s shape, magnetic field morphology, or viewing angle. Even if we were to overlook this issue, a second problem arises. For projection angles where $\rm{B_{POS}}$ is best measured (i.e., when the field lies mostly in the plane of the sky), the column density we observe is perpendicular to the field lines. This means that we are not measuring the mass along the flux tubes, which is the quantity that enters the definition of the mass-to-flux ratio, but rather the mass across them. 

Given these limitations, one might still be tempted to estimate $\mu_{\rm obs}$ by inferring a ``reasonable'' total magnetic field strength from a single component by invoking statistically motivated corrections about the field's orientation. For instance, assuming a large ensemble of clouds with randomly oriented magnetic fields relative to the LOS, one can show that, on average, $\rm{\overline{B_{LOS}}} = (1/2)~\rm{\overline{B_{total}}}$ (e.g., \citealt{1982ApJ...260L..23H}). In this context, one might infer $\rm{B_{total}}$ from $\rm{B_{LOS}}$ (or $\rm{B_{POS}}$) and estimate the ``total mass-to-flux ratio''. However, this approach conflates ensemble averages with individual measurements, which can lead to misleading conclusions. It is analogous to expecting to find exactly 2.54 people in a particular household simply because that is the average number of people per household. Thus, such ``corrections'' can introduce significant errors when applied to a single cloud, and in practice, many studies continue to report ``mass-to-flux ratios'' based directly on $\rm{B_{POS}}$ measurements or similarly oversimplified assumptions.

Here, we make use of a 3D numerical simulation to study projection angle effects on estimates of the mass-to-flux ratio using the magnetic field-gas column density relation \citep{1983AdSpR...2l..71M, 1987ASIC..210..453M}. Our paper is organized as follows: In Sect.~\ref{methods} we outline the basic details of the modeled cloud, and the methodology followed for producing idealized mock observations. In Sect.~\ref{Results} we present our results on the estimated mass-to-flux ratio using both LOS and POS estimates of the magnetic field. We summarize our findings in Sect.~\ref{discuss}.


\section{Methodology}\label{methods}

To study the mass-to-flux ratio, we use the simulation of a collapsing molecular cloud first presented in \cite{2025A&A...700A.152T}. \cite{2025A&A...700A.152T} modeled the evolution of a $\sim$240 M\textsubscript{\(\odot\)} supercritical ($\mu/\mu_{crit} \sim 1.7$) sub-Alfv\'enic ($\mathcal{M_\mathrm{A}}\sim0.85$) supersonic ($\mathcal{M_\mathrm{s}}\sim3$) isothermal molecular cloud under nonideal MHD conditions and decaying turbulence. Together with the dynamical evolution, they also modeled the chemical evolution of the cloud by employing a nonequilibrium network consisting of 115 species, including gas-phase and ice-mantle species (for more details on the chemical network we refer to \citealt{2016MNRAS.458..789T, 2022MNRAS.510.4420T}). The abundances of the charged species in the network were then used to calculate the resistivities of the cloud.

The temperature and initial density of the cloud were 10~K and 500 $\rm{cm^{-3}}$, respectively. The magnetic field was initially along the $z$ axis of the cloud and its strength was 10 $\rm{\mu G}$. Turbulence was initiated using the publicly available \texttt{TurbGen} code \citep{2022ascl.soft04001F, 2010A&A...512A..81F} assuming a power law velocity ($v$) spectrum of $dv^2/dk \propto k^{-2}$, where $k$ is the wavenumber. The dimensions of the cloud were 2 pc in each direction. The simulation was performed using a modified version of the \textsc{FLASH} astrophysical code \citep{2000ApJS..131..273F, 2008ASPC..385..145D} on a $64^3$ grid with two levels of adaptive mesh refinement and open boundary conditions.

In order to study the magnetic field-gas column density relation we used the following approach. First, we randomly selected a snapshot of the simulation at a time of $t$ $\in$ [0.5, 0.75, 1., 1.25, 1.44] $\times$ $t_{ff}$, where $t_{ff}$ is the free-fall time. We then randomly projected the cloud at an inclination angle of $\gamma$ $\in$ [0$^\circ$, 22.5$^\circ$, 45$^\circ$, 67.5$^\circ$, 90$^\circ$]. The inclination angle corresponds to the polar angle (i.e., the angle between the LOS and the initial magnetic field direction, see angle $\phi$ in Figure 1 in \citealt{2025A&A...696A..35T}). For each time and projection angle we computed the column density of the cloud, the 3D POS and LOS magnetic field components of the cloud, and the 3D number-density distributions of species $\rm{OH}$, $\rm{CN}$, and $\rm{CO}$ along the LOS, as these were computed from our chemical network. The reasoning behind selecting $\rm{OH}$ and $\rm{CN}$ is because they are used to observe the Zeeman effect, while $\rm{CO}$ is commonly used in combination with polarization measurements to probe the magnetic field. We then placed a circular beam at a random location within 0.5 pc from the center of the cloud. The beam radius was randomly sampled under the condition that the resulting region remains entirely within the cloud boundaries. The minimum beam radius was set to one pixel. Finally, we compute the values of the POS and LOS components of the magnetic field inside the beam. For $\rm{B_{POS}}$ estimates, a one-pixel beam would, in observational terms, imply deriving the field strength from a single polarization segment which is unphysical. However, in our mock framework, the $\rm{B_{POS}}$ values were not inferred from polarization angles or dispersion, but rather directly computed from the simulation as described below.

To compute the LOS component of the field we consider the average of the 3D values of the LOS component weighted by the molecular number density\footnote{The total weight per cell along the LOS is the product of molecular number density and the distance traversed inside that cell.}. Considering a weighted average of the field for our mock ``Zeeman measurements'' is a very well justified choice since the observed signal depends on the magnetic field's direction, and as such, contributions from regions of opposite polarity within the same resolution element cancel out (e.g., \citealt{1998A&A...329..319S}). As a result, only the net (i.e., uncanceled) component of the LOS magnetic field contributes to the Zeeman signal. This implies that for a perfectly axisymmetric cloud with an hourglass magnetic-field morphology seen edge on, the magnetic field measured in Zeeman observations will be zero. Motivated by observations \citep{2019FrASS...6...66C}, we also follow a probabilistic approach to decide which of the two species ($\rm{OH}$ or $\rm{CN}$) will be used to weigh the magnetic field. First, we compute the mean value of the column density inside the beam. We then define two Gaussian functions centered at characteristic column densities at which each species is used to observe the Zeeman effect. We set the means of the Gaussians to be $8\times10^{21}~\rm{cm^{-2}}$ and $4\times10^{22}~\rm{cm^{-2}}$ for $\rm{OH}$ and $\rm{CN}$, respectively. The standard deviations of the Gaussians are equal to 0.2 in $\rm{log_{10}}$ space. In practice, this means that regions of the cloud with column densities near $8\times10^{21}~\rm{cm^{-2}}$ are more likely to be observed (i.e., weighted) with $\rm{OH}$, whereas regions with column densities near $4\times10^{22}~\rm{cm^{-2}}$ are more likely to be observed with $\rm{CN}$. Finally, the process described above was repeated 5$\times 10^3$ times.

For our mock ``measurements'' of the POS component of the magnetic field we follow a different approach. In a recent study, \cite{2025A&A...700A.256P} systematically investigated POS measurements of the magnetic field based on the the Davis-Chandrasekhar-Fermi (DCF) \citep{1951Phys.Rev....81...890, 1953ApJ...118..113C} and Skalidis-Tassis (ST) \citep{2021A&A...647A.186S} methods. In that work, the authors analyzed an ideal MHD simulation with nonequilibrium chemical modeling, and generated synthetic polarization maps using the method of \cite{2019MNRAS.490.2760K}, incorporating a spatially varying polarization efficiency based on the local (3D) visual extinction, and density of each cell. The simulation was additionally post-processed with a non-local thermodynamic equilibrium radiative transfer code to produce synthetic spectra (i.e., the $J = 1\rightarrow0$ \& $J = 2\rightarrow1$ transitions of $\rm{CO}$). They then extracted the parameters required to estimate the strength of the POS component of the magnetic field ($\rm{\rho}$, $\delta v$, $\delta \theta$) using multiple approaches, under various projection angles. Their analysis showed that both the DCF and ST methods trace the median of the molecular-species–weighted (i.e., the $\rm{CO}$-number-density-weighted) POS component of the magnetic field (see their Fig. 9). We therefore apply the same weighting scheme to compute the $\rm{B_{POS}}$ value, and assign the resulting weighted average to the corresponding mean column density within the beam\footnote{A mass-weighted $\rm{B_{POS}}$ would overemphasize contributions from dense regions along the LOS, which are, however, typically associated with depolarization. Using mass weighting yields higher $\rm{B_{POS}}$ values, particularly at later evolutionary stages, when the mass spread among cells along the LOS becomes larger. This, in turn, yields lower mass-to-flux ratios, which misleadingly suggest that the mass-to-flux ratio decreases over time.}. An implicit assumption in this methodology is that the underlying relation between the true $\rm{B_{POS}}$ and the observed quantities can be meaningfully applied down to the resolution limit. Similar to the mock Zeeman measurements, this process was repeated 5$\times 10^3$ times.

\section{Results}\label{Results}

\begin{figure*}
\def\arraystretch{0.0}
\begin{tabular}{l}
\includegraphics[width=2.\columnwidth, clip]{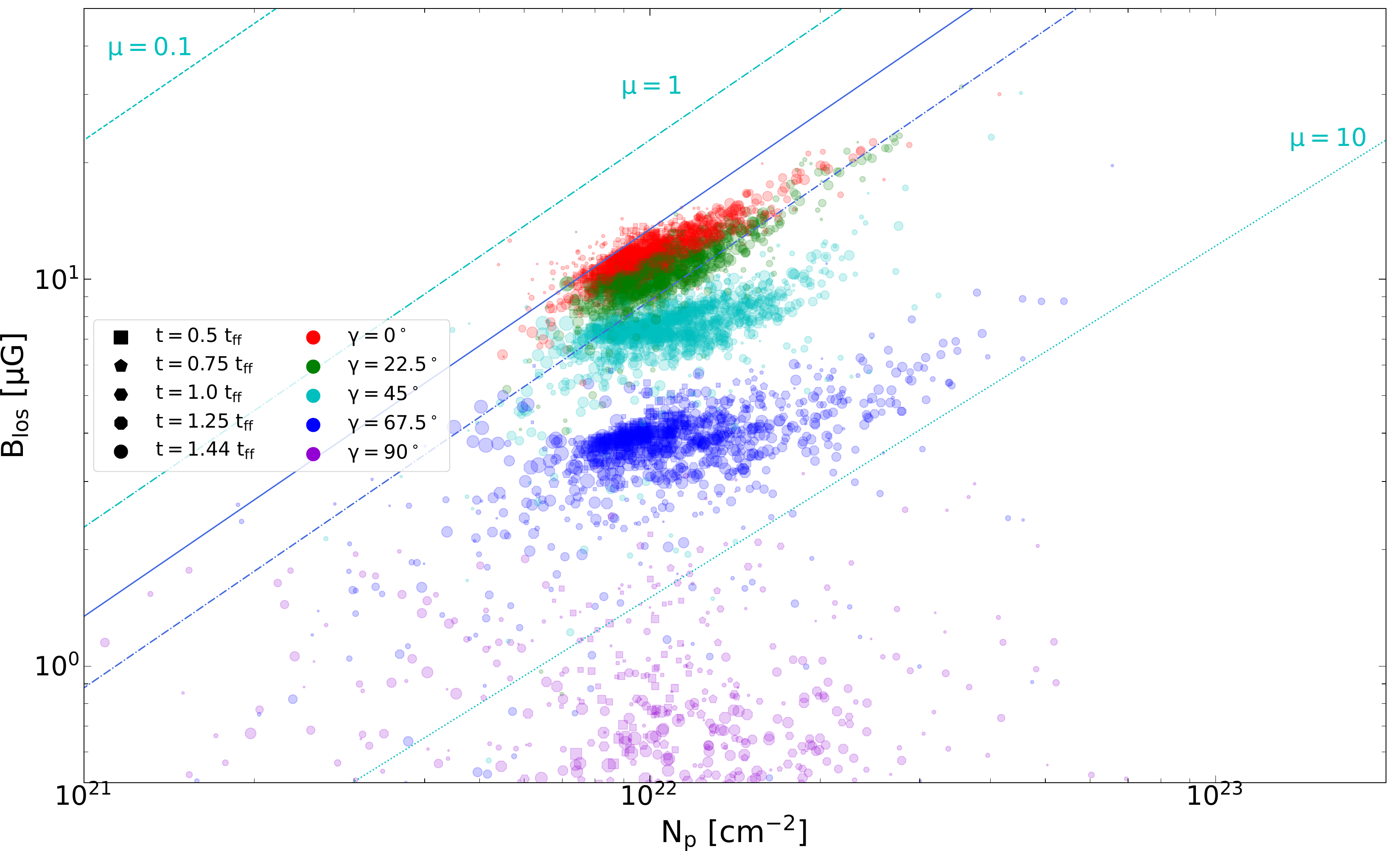} \\
\includegraphics[width=2.\columnwidth, clip]{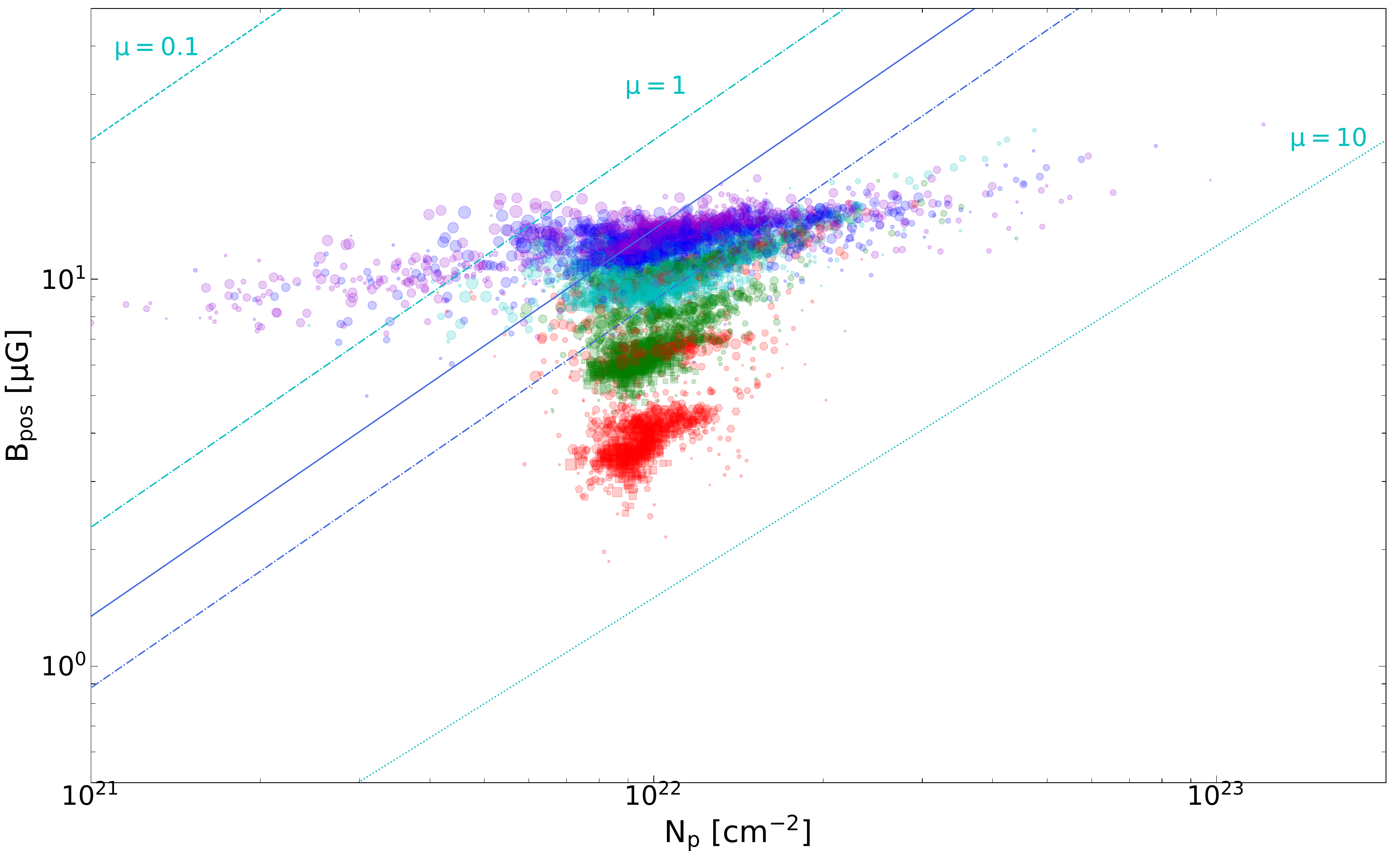}
\end{tabular}
\caption{Magnetic field column density relation from our simulation considering the LOS and POS components of the field (upper and lower panels, respectively). The cyan dashed, dash-dotted and dotted lines mark constant mass-to-flux ratios of $\mu/\mu_{crit}$ of 0.1, 1, and 10, respectively. The thin blue solid line indicates the initial value of the mass-to-flux ratio in the cloud and the blue dashed line marks the maximum value of the mass-to-flux ratio at a time of 1.44$\times~t_{ff}$. Colored points correspond to different projection angles, different symbols correspond to different times when the cloud is ``observed'', and the size of the symbols is proportional to the size of the beam used to observe different regions of the cloud (see Sect.~\ref{methods}).
\label{BCDrel}}
\end{figure*}

In Fig.~\ref{BCDrel} we present our results on the magnetic field-gas column density relation considering the LOS (upper panel) and POS (lower panel) components of the magnetic field. The cyan dashed, dash-dotted and dotted lines indicate mass-to-flux ratios of $\mu/\mu_{crit}$ of 0.1, 1, and 10, respectively, and the thin blue solid line marks the initial value of the mass-to-flux ratio in the modeled cloud. We note here, that the value of the true mass-to-flux ratio at a time of 1.44 $\times$ $t_{ff}$, when measured by following magnetic field lines, is up to 2.6 the critical value for collapse in the central regions of the cloud (see Figs. 5 and 6 from \citealt{2025A&A...700A.152T}). This value of the mass-to-flux ratio is marked with the blue dashed line. Colored points correspond to different inclination angles under which the cloud is observed. Specifically, with the red, green, cyan, blue and magenta points we show, respectively, our results for inclination angles of $\gamma = 0^\circ, 22.5^\circ, 45^\circ, 67.5^\circ$, and $90^\circ$. With the square, pentagon, hexagon, octagon, and circle symbols we show our results when we observe the modeled cloud at a time of 0.5, 0.75, 1., 1.25, 1.44 $\times$ $t_{ff}$, respectively. Finally, the size of the symbols corresponds to the size of the circular beam.

The projection effects discussed in Sect.~\ref{intro}, which influence Zeeman measurements, are clearly evident in the upper panel of Fig.~\ref{BCDrel}. The observed mass-to-flux ratio $\mu_{\mathrm{obs}}$ aligns with or falls within the range of the true initial and maximum values of the mass-to-flux ratio only when the cloud is viewed approximately face-on (i.e., for $\gamma \leq 22.5^\circ$). However, we note that even in these favorable inclinations, $\mu_{\mathrm{obs}}$ does not necessarily spatially exactly correspond to the true mass-to-flux ratio. This is due to the fact that, in observations, we cannot track the 3D structure of magnetic flux tubes \citep{2009MNRAS.400L..15M, 2025A&A...700A.152T}. Nonetheless, for these inclination angles, Zeeman measurements can reasonably probe the dynamical significance of the magnetic field with respect to the cloud's self-gravity. For higher inclination angles, the measured magnetization of the cloud increasingly deviates from the true mass-to-flux ratio (see cyan, blue and magnenta points in the upper panel of Fig.~\ref{BCDrel}). This discrepancy becomes especially pronounced when the cloud is observed ``edge on'' (i.e., the mean component of the filed lies on the POS). For these inclination angles the scatter in the magnetic field gas density relation increases significantly and the slope of the relation systematically changes to shallower values.

Estimates of the mass-to-flux ratio derived from polarization measurements (lower panel of Fig.~\ref{BCDrel}) are affected by a different set of systematics. At first glance, the points corresponding to different inclination angles appear to lie closer to the true initial and maximum mass-to-flux ratios from the simulation for all projection angles. However, this apparent better agreement is simply due to the fact that, polarization measurement, are not sensitive to the polarity of the field along the LOS, but rather to its tangling and orientation with respect to the POS. Additionally, upon closer inspection, two particularly problematic issues become clearly evident. First, for all inclination angles, the slope of the points in the magnetic field–column density plane deviates significantly from that expected based on the true variations in the mass-to-flux ratio. This means that, although some data points may fall between the lines marking the initial and maximum values of $\mu$, they do so along a trajectory that is inconsistent with the physical evolution of the cloud. In other words, the alignment with the expected $\mu$ range is merely coincidental and not diagnostic. Therefore, for the same inclination angle, POS measurements of the field can result in any value for the mass-to-flux ratio, from subcritical to highly supercritical. Second, and equally concerning, is that for certain projection angles (e.g., $\gamma \leq 45^\circ$), the inferred mass-to-flux ratio appears to decrease as the cloud evolves in time, contrary to the physical evolution of the cloud. For instance, square symbols (representing $0.5\times~t_{ff}$) exhibit higher mass-to-flux ratios than circular markers (representing the final evolutionary stage of the simulation). This inverse behavior is unphysical and highlights the potential for polarization-based measurements to yield misleading interpretations about the role of magnetic fields in cloud evolution. This issue is not merely a subtle caveat. If commonly used observational diagnostics yield invalid conclusions because they are fundamentally flawed and/or biased, this discrepancy must be clearly identified to avoid drawing and perpetuating incorrect conclusions in the literature (see e.g., the debate between \citealt{2009ApJ...692..844C} and \citealt{2010MNRAS.409..801M} on mass-to-flux ratio measurements and the assumption of straight-parallel field lines).

Finally, we also note that the apparent clustering of points around the correct values of the mass-to-flux ratio for DCF and/or ST measurements is largely driven by the idealized conditions (e.g., uniform initial density and square-grid geometry) of the simulation. This has the following effect. Suppose that we observe the modeled cloud at a time of t = 0 (when the cloud still has uniform density) such that the magnetic field lies entirely on the POS. With DCF we would see $\rm{B_{POS}} = \rm{B_{total}} = \rm{B_{initial}}$. In this configuration we will observe the mass perpendicular to the flux tubes. However, because of the initially uniform density and square geometry, the mass perpendicular to the flux tubes is equal to the mass parallel to the flux tubes. As such, we would ``derive'' a reasonable value for the mass-to-flux ratio. This apparent and coincidental agreement, however, comes from idealized conditions and not from the physical validity of the method.

\begin{figure}
\includegraphics[width=1.\columnwidth, clip]{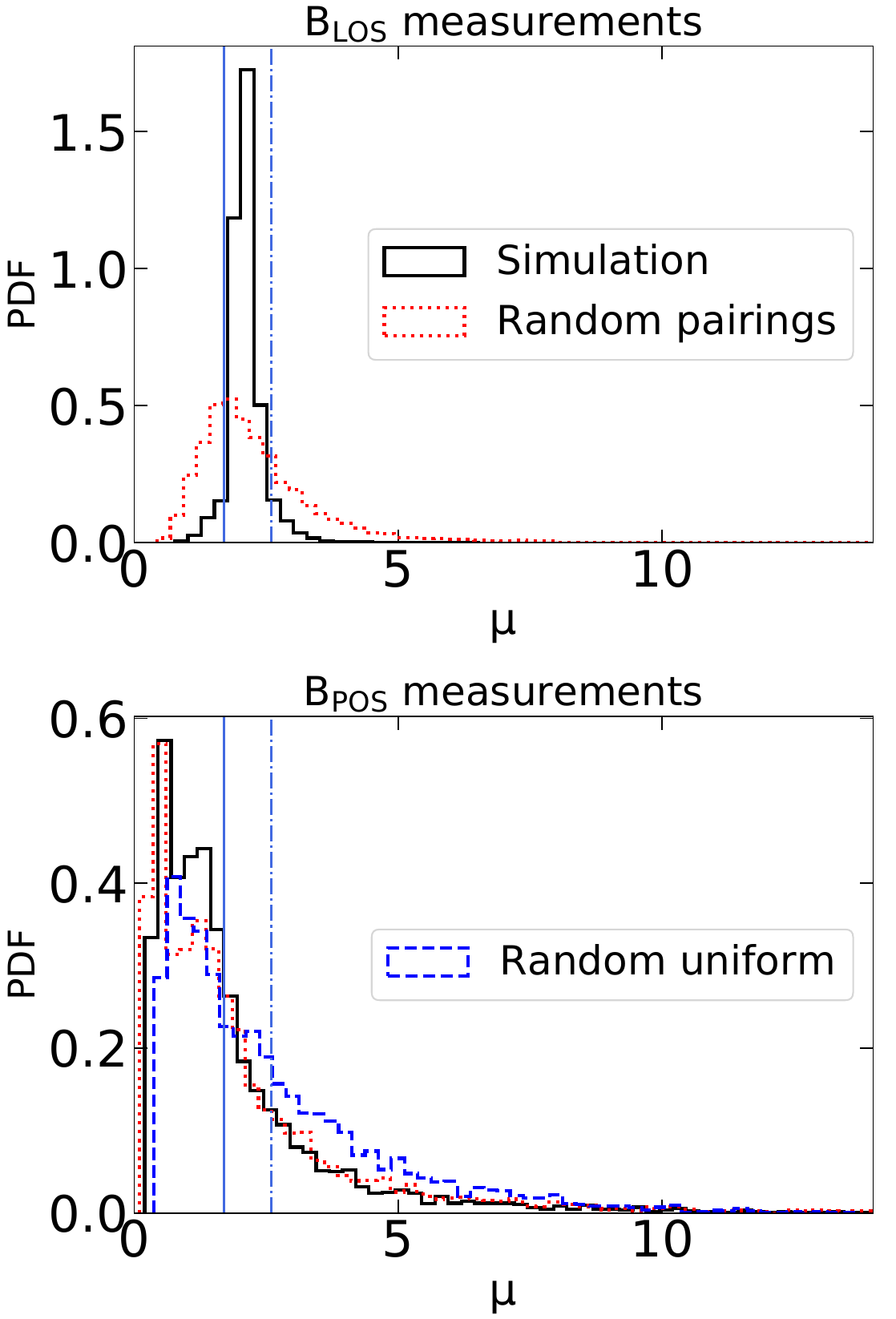}
\caption{PDFs of the mass-to-flux ratio based on the LOS (upper panel) and POS (lower panel) components of the magnetic field for a time of $1.44\times t_{ff}$, when the cloud has evolved to a more realistic configuration. For each case, we adopt the inclination angle that maximizes the respective component ($\gamma = 0^\circ$ for $\rm{B_{LOS}}$ and $\gamma = 90^\circ$ for $\rm{B_{POS}}$). Black curves correspond to the results from the simulation. Red dotted lines represent PDFs obtained when column densities and magnetic field strengths from different regions of the cloud (and with different beam sizes) are randomly paired. The blue dashed line shows a PDF of mass-to-flux ratios obtained by considering independent uniform random distributions for $\rm{B_{POS}}$ and $\rm{log_{10}N_p}$. The blue solid and dash-dotted vertical lines indicate the initial and maximum values of the true $\mu$, respectively. At late times, only $\sim$17\% of $\rm{B_{POS}}$ estimates fall within the range defined by the true mass-to-flux ratio, while most values suggest a substantially subcritical cloud. The similarity between the $\rm{B_{POS}}$ distribution from the simulation, and that obtained from random pairings, indicates that such measurements do not preserve the underlying correlation between magnetic field strength and column density.
\label{hists}}
\end{figure}

To further highlight these points, we show in Fig.~\ref{hists} probability density functions (PDFs) of the mass-to-flux ratio estimated from $\rm{B_{LOS}}$ (upper panel) and $\rm{B_{POS}}$ (lower panel) at a time of $t = 1.44 \times t_{\rm ff}$, at which stage, the cloud has evolved to a more realistic configuration. For each case ($\rm{B_{LOS}}$ and $\rm{B_{POS}}$), we adopt the inclination angle for which the corresponding component is optimally observed (i.e., $\gamma = 0^\circ$ for $\rm{B_{LOS}}$, $\gamma = 90^\circ$ for $\rm{B_{POS}}$). We performed $5 \times 10^3$ independent measurements by varying the beam location and size. The results obtained from the simulation are shown as black curves whereas the red dotted lines show the PDFs of the mass-to-flux ratio obtained by randomly pairing different column density and magnetic field measurements from different parts of the cloud. The reasoning behind performing random pairings is to break any existing correlations (if any) between the magnetic field and the column density. Finally, the blue dashed line depicts the PDF obtained by simply considering independent uniform random distributions for $\rm{B_{POS}}$ and $\rm{log_{10}N_p}$ within the ranges [5, 20] $\rm{\mu G}$ (i.e., a factor of two above and below the initial value of the magnetic field in the simulation), and [21.5, 22.5] $\rm{cm^{-2}}$ (i.e., reasonable limits based on the column density of the cloud), respectively. We therefore test the null hypothesis, if estimates of the mass-to-flux ratio based on $\rm{B_{POS}}$ have any diagnostic power beyond randomness. As in Fig.~\ref{BCDrel}, the initial and maximum values of the true mass-to-flux ratio are marked as blue solid and dash-dotted lines, respectively.

Two features stand out from Fig.~\ref{hists}. For $\rm{B_{LOS}}$, the vast majority of measurements (88\%) fall between the initial and maximum true values of the mass-to-flux ratio. For $\rm{B_{POS}}$, the picture is very different. Only $\sim$17\% of mass-to-flux ratio estimates fall within the true bounds\footnote{For other inclination angles, the fraction of estimates within the true boundaries increases to 22\% ($\gamma = 67.5^\circ$) and 42\% ($\gamma = 45^\circ$).}. Additionally, $\sim$35\% of the estimates are more than a factor of two away from both limits. For this particular cloud and inclination, $\rm{B_{POS}}$ estimates would likely result in the cloud appearing subcritical. Specifically, the mode of the PDF of observed mass-to-flux ratios is $\sim 0.7$, more than a factor of three away from the ``average'' true value. Secondly, when column densities and field strengths from different sightlines and beam sizes are randomly paired, the PDF of observed mass-to-flux ratio measurements based on $\rm{B_{LOS}}$ is much broader, indicating a clear loss of information. The same is not true for estimates based on $\rm{B_{POS}}$ where the two PDFs (directly from the simulation and random pairings) are similar. We note that the same is true for other inclination angles. This reflects that such measurements do not encode the correlation between the magnetic field and the column density in the first place. Finally, in the PDF of mass-to-flux ratios obtained by considering independent uniform random distributions for $\rm{B_{POS}}$ and $\rm{log_{10}N_p}$, 19\% of the estimates fall within the true boundaries, essentially indistinguishable from the DCF case. This test demonstrates that DCF estimates of the mass-to-flux ratio perform no better than this purely random baseline.


\section{Discussion and conclusions}\label{discuss}

We employed the nonideal MHD chemo-dynamical simulation of a supercritical transAlfv\'enic supersonic collapsing molecular cloud from \cite{2025A&A...700A.152T}, and the magnetic field-gas column density relation to investigate biases in observed values of mass-to-flux ratio. Our analysis shows that projection effects play a critical role in shaping the inferred values. When the mean magnetic field lies predominantly in the plane of the sky, the intrinsic magnetization of the cloud (relative to its self-gravity) can be overestimated by more than an order of magnitude. As such, Zeeman-based estimates of the mass-to-flux ratio should generally be regarded as upper limits.

For mass-to-flux ratio estimates based on the POS component of the magnetic field, the situation is more problematic, albeit for different reasons. Firstly, defining a mass-to-flux ratio based on such measurements is physically and mathematically erroneous, as the flux of $\vec{\rm{B_{POS}}}$ through the POS is, ipso facto, zero. Secondly, the trend traced by the points in the magnetic field–column density plane does not reflect the true evolution of the mass-to-flux ratio. This misalignment implies that any apparent agreement with true values is largely coincidental. Third, in configurations where we can most reliably measure $\rm{B_{POS}}$, we measure the mass perpendicular, rather than along, the magnetic flux tubes. That is, the wrong mass for computing the mass-to-flux ratio. Finally, such flawed estimates of the mass-to-flux ratio can lead to additional misleading conclusions where already collapsed clouds appear to be more magnetically supported than their earlier stages.

One potential limitation of our analysis is the omission of radiative transfer effects in the mock derivation of the LOS magnetic field component. Such effects, along with pitfalls of the analysis technique, have been thoroughly examined in \citet{2017A&A...601A..90B}. However, the impact of the inclination angle in Zeeman measurements of the mass-to-flux ratio is so severe that the qualitative conclusions are unlikely to change. If anything, including radiative transfer and related observational complexities would likely exacerbate any discrepancies rather than mitigate them. \citet{2023MNRAS.521.5604H} also studied the magnetic field-column density relation in mesh-free galactic zoom-in simulations. They argued that H\textsc{I}-based estimates of column density may significantly underestimate the true total hydrogen column, leading to clouds appearing subcritical even if they are not. However, even upon adopting their correction factors for the missing mass (see their Fig. 9), the implied changes in column density are not sufficient to cross the supercritical line, let alone reach the higher values of the mass-to-flux ratios implied by their Fig. 8. For instance, reaching the $\mu=5$ line from a column density of $1 \times 10^{19}~\mathrm{cm}^{-2}$ would require a correction factor of $\sim$400, whereas their Fig. 9 suggests that the missing mass is smaller by $\sim$one order of magnitude. Therefore, the discrepancy they observe may instead arise from a boundary effect. Near the simulation edge, a sightline may intersect only a small number of particles, potentially underestimating the column density while still assigning a reasonable strength for the magnetic field. This combination would lead to artificially low inferred mass-to-flux ratios. Regardless, the column densities probed in the present study correspond to dense regions within molecular clouds, where observational estimates of the column density would typically rely on molecular line emission and/or thermal dust continuum rather than H\textsc{I}. As such, any of the physical biases discussed in \citet{2023MNRAS.521.5604H} are not expected to affect the densities or scales examined here.

\begin{acknowledgements}

We thank K. Tassis, and R. Skalidis for stimulating discussions and comments. We also thank the anonymous referees for questions and suggestions which helped enhance the clarity and quality of our manuscript. A. Tritsis acknowledges support by the Ambizione grant no. PZ00P2\_202199 of the Swiss National Science Foundation (SNSF), and the MERAC Foundation. The software used in this work was in part developed by the DOE NNSA-ASC OASCR Flash Center at the University of Chicago. This research was enabled in part by support provided by SHARCNET (Shared Hierarchical Academic Research Computing Network) and Compute/Calcul Canada and the Digital Research Alliance of Canada. We also acknowledge use of the following software: \textsc{Matplotlib} \citep{2007ComputSciEng.9.3}, \textsc{Numpy} \citep{2020Nat.585..357}, \textsc{Scipy} \citep{2020NatMe..17..261V} and the \textsc{yt} analysis toolkit \citep{2011ApJS..192....9T}.

\end{acknowledgements}

%
%

\begin{appendix}\onecolumn

\section{Column density maps}\label{cds}

\begin{figure}
\includegraphics[width=1.\columnwidth, clip]{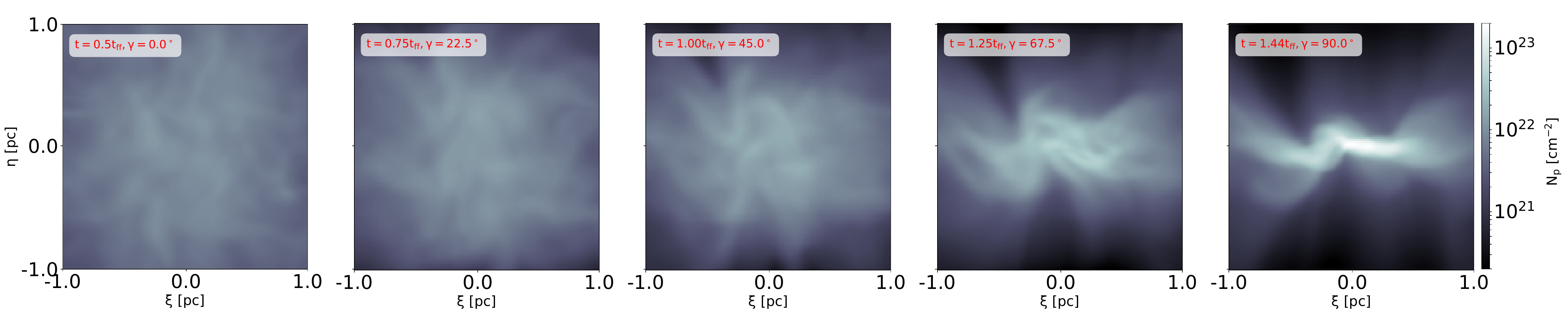}
\caption{Column density maps of the simulation used, at different times, and inclination angles. From left to right, we present the following combinations $[0.5\times t_{ff}, ~0^\circ]$, $[0.75\times t_{ff}, ~22.5^\circ]$, $[1.0\times t_{ff}, ~45^\circ]$, $[1.25\times t_{ff}, ~67.5^\circ]$, $[1.44\times t_{ff}, ~90^\circ]$.
\label{cdPlots}}
\end{figure}

In this appendix, we present column density maps from the simulation used throughout this work, to illustrate how the projected morphology evolves with time and inclination. In Fig.~\ref{cdPlots}, we show the column density for five representative combinations of evolutionary time and inclination angle. From left to right, we present our results for $[0.5\times t_{ff}, ~0^\circ]$, $[0.75\times t_{ff}, ~22.5^\circ]$, $[1.0\times t_{ff}, ~45^\circ]$, $[1.25\times t_{ff}, ~67.5^\circ]$, $[1.44\times t_{ff}, ~90^\circ]$. For comparison, column density maps from a similar simulation at a fixed evolutionary time and varying projection angles are shown in Fig. 2 of \cite{2025A&A...696A..35T}, while the spatial coordinates $\eta$ and $\xi$ on the POS are explained in their Fig. 1. These snapshots provide a visual reference for how changes in the viewing angle and dynamical evolution affect the appearance of the cloud as projected on the POS. As expected, the largest degree of spatial variation in column density appears at $[1.44\times t_{ff}, ~90^\circ]$, whereas the most uniform distribution occurs at $[0.5\times t_{ff}, ~0^\circ]$.

\end{appendix}

\end{document}